\documentclass[prl,twocolumn,twoside,showpacs,superscriptaddress]{revtex4}
\usepackage{graphics,amssymb,amsmath,epsf}
\epsfclipon

\usepackage{xspace}

\newcommand{\be}{\begin{equation}}
\newcommand{\ee}{\end{equation}}

\newcommand{\ga}{\tilde{g}_{11}}
\newcommand{\gb}{\tilde{g}_{02}}

\newcommand{\p}{\partial} 
\newcommand{\tpsi}{\tilde{\psi}}
\newcommand{\vx}{\vec{x}}
\newcommand{\vv}{\vec{v}}
\newcommand{\bq}{{\bf q}}
\newcommand{\bp}{{\bf p}}

\newcommand{\tq}{\tilde{q}}
\newcommand{\tp}{\tilde{p}}
\newcommand{\tu}{\tilde{u}}
\newcommand{\tQ}{\tilde{Q}}
\newcommand{\tP}{\tilde{P}}
\newcommand{\tU}{\tilde{U}}
\newcommand{\tH}{\tilde{H}}

\newcommand{\dd}{\text{\tiny $D$}}
\newcommand{\etan}{\eta_{\nu}}
\newcommand{\etad}{\eta_{\dd}}
 
\newcommand{\tih}{\tilde{h}}
\newcommand{\tj}{\tilde{j}}

\begin{document}

\title{Non-perturbative renormalization group for the Kardar-Parisi-Zhang equation}

\author{L\'eonie Canet}
\affiliation{LPMMC, CNRS UMR 5493, Universit\'e Joseph Fourier, 38042 Grenoble, France}
\author{Hugues Chat\'e} 
\affiliation{Service de Physique de l'Etat Condens\'e, CEA-Saclay, 91191 Gif-sur-Yvette, France}
\author{Bertrand Delamotte}
\affiliation{LPTMC, CNRS UMR 7600, Universit\'e Pierre et Marie Curie, 75252 Paris, France}
\author{Nicol\'as Wschebor}
\affiliation{Instituto de F\'{\i}sica, Faculdad de Ingenier\'{\i}a, 
Universidad de la Rep\'ublica, 11000 Montevideo, Uruguay}

\begin{abstract}
We present a simple approximation of the
non-perturbative renormalization group designed for the Kardar-Parisi-Zhang equation
and show that it yields the correct phase diagram, including the strong-coupling phase
with reasonable scaling exponent values in physical dimensions.
We find indications of a possible qualitative change of behavior around $d=4$.
We discuss how our approach can be systematically improved.
\end{abstract}
\maketitle

The Kardar-Parisi-Zhang (KPZ) equation \cite{kardar86}:
\begin{equation}
\frac{\p h(\vx,t)}{\p t} = \nu\,\nabla^2 h(\vx,t) \, + 
\,\frac{\lambda}{2}\,\big(\nabla h(\vx,t)\big)^2 \,+\,\eta(\vx,t)
\label{eqkpz}
\end{equation}
where $\eta$ is an uncorrelated Gaussian noise with zero mean and variance
$\big\langle \eta(\vx,t)\eta(\vx',t')\big\rangle = 2\,D\,\delta^d(\vx-\vx')\,\delta(t-t')$
 is maybe the simplest nonlinear Langevin equation showing non-trivial behavior \cite{halpin95}. 
Under form (\ref{eqkpz})\, 
it describes the kinetic roughening of a $d$-dimensional interface \cite{kardar86}, 
but it is now recognized as representing an extremely large class of non-equilibrium or
disordered systems \cite{halpin95}. 
A large body of works at all levels have led to a general agreement about the phase diagram
of the KPZ equation, usually described in terms of the dynamical scaling properties of 
$h(\vec{x},t)$,
 characterized by the two-point correlation function 
$C(|\vx-\vx'|,t-t') \equiv \langle [h(\vx,t) - h(\vx',t')]^2\rangle$. In particular,
 at large scales, 
$C$ shows the scaling form $C(L,\tau) = L^{2\chi}\,f(\tau/L^z)$ with the two exponents
related via $z+\chi=2$. 
For $d\le2$,  the nonlinear term is relevant and the interface always roughens 
($\chi>0$).
For $2<d<4$, two phases exist depending on the value of the effective coupling constant
$g=\lambda^2 D/\nu^3$: in the strong-coupling regime, rough interfaces are again observed,
while for small $g$ values the nonlinear term is irrelevant 
and the interface is smooth ($\chi=0, z=2$). 

Some important points remain controversial:
For $d\ge4$, numerics \cite{tang92} and real-space calculations 
\cite{castellano98} still show
the existence of both a rough and a smooth phase, while
some theoretical approaches \cite{lassig97,colaiori01,fogedby05} 
argue that $d=4$ is some kind of upper critical dimension beyond which $z=2$.
(For a recent discussion, see \cite{Katzav1}.)
Thanks to an incidental extra symmetry, $\chi=\frac{1}{2}$ is known in $d=1$ \cite{kardar87},
but no exact results are available in higher dimensions.
This unsatisfactory situation is largely due to the fact that 
the strong coupling phase 
of the KPZ equation has remained out of reach of controlled analytical approaches. 
In particular, standard perturbation expansions fail {\it at all order} to find a 
strong coupling fixed point \cite{wiese98}. Some non-perturbative approaches have  
been deviced, such as the mode coupling approximation 
\cite{beijeren85,colaiori01}, the self-consistent expansion \cite{SCE}, or the weak noise 
scheme \cite{fogedby05}, but they are difficult to improve in a systematic way.
In this context, the non-perturbative renormalization group (NPRG) \cite{berges02} appears 
promising, in particular since it is able to deal with
perturbatively unaccessible fixed points even in out-of-equilibrium problems
\cite{canet}.
In this Letter, we present a simple approximation of the
NPRG designed for the KPZ equation
and show that it yields the correct phase diagram, including the strong-coupling phase
with reasonable scaling exponent values in physical dimensions.
In particular the strong-coupling fixed point in $d\ge2$ is 
{\it genuinely non-perturbative} (hence essentially out of reach of perturbative RG) 
and {\it fully attractive} --- which roots the existence of generic scaling. 
We find indications that a qualitative change of behavior occurs around $d=4$,
but at the minimal order presented here, our approximation does not resolve the 
puzzle of the existence of an upper critical dimension. 
We finally discuss how our approach 
can be systematically improved.

We first recall how the KPZ problem can be cast into a field theory \cite{janssen76}.
Introducing a response field $\tih$ and sources $(j,\tj)$,
the generating functional reads:
\begin{eqnarray}
{\cal Z}[j,\tj] \!\! &=& \!\!\!\int {\cal D}[h,i \tih]\, 
\exp\left(-{\cal S}[h,\tih] +  \int_{\bf x} (j\,h+\tj\tih) \right)\label{Z}\\
{\cal S}[h,\tih]  \!\! &=& \!\!\! \int_{\bf x}  \left\{ \tih\left(\p_t h -\nu \,\nabla^2 h - 
\frac{\lambda}{2}\,({\nabla} h)^2 \right) - D\, \tih^2  \right\}
\label{action}
\end{eqnarray} 
where ${\bf x}=(\vx,t)$.
The symmetries of the KPZ equation are three-fold:
 (i) invariance of ${\cal Z}$ under the Galilean transformation
$
{\cal T}_{\rm G}=\{h({\bf x})\to h(\vx+ \lambda\vv t,t)+ \vv.\vx, \tih({\bf x})\to \tih(\vx+\lambda\vv t,t)\};
$
(ii) invariance of the combination ${\cal S} -\int \tih \p_t h$ under the 
``time-gauged'' symmetry
${\cal T}_{\rm TG} =\{h({\bf x})\to h({\bf x})+f(t)\}$ 
where $f(t)$ is an arbitrary function of time;
(iii) in $d=1$,  additional time-reversal  invariance
$
{\cal T}_{\rm TR}=\{h(t)\to -h({-}t), \tih(t)\to \tih({-}t)+\frac{\nu}{2D}\, \nabla^2 h({-}t)\}
$.

The NPRG builds a one-parameter family of models indexed by a scale $k$  
such that fluctuations are smoothly included as $k$ is lowered 
from the  microscopic scale $\Lambda$, where no fluctuations are taken 
into account, to $k=0$ where they have all been summed over \cite{berges02}. 
To this aim, we add, 
to the original action ${\cal S}$,
a momentum-dependent mass-like term $\Delta {\cal S}_k$
which preserves the symmetries:
\begin{equation}
\hspace{-2pt} \Delta {\cal S}_k \!=\!\frac{1}{2}\! \int_{\bf q}\!  h_i\, [R_k]_{ij}\, h_j  \;\;;\;
R_k \!=\! r(\frac{q^2}{k^2})
\left(\!\! \begin{array}{cc}
0& \nu_k q^2\\
\nu_k q^2 & -2 D_k
\end{array}\!\!\right)
\label{deltask}
\end{equation}
where ${\bf q}=(\vec{q},\omega)$, $q=\|\vec{q}\|$, $i,j\in\{1,2\}$, $h_1=h,h_2=\tih$, and 
summation over repeated indices is implicit.
With the choice $r(x)=\alpha/(\exp(x) -1)$ where $\alpha$ is a parameter, 
the fluctuation modes $h_i(q \gtrsim k)$ are unaffected by $\Delta {\cal S}_k$,
while the others ($h_i(q\lesssim k)$) are essentially frozen.
The effective action $\Gamma_k[\psi,\tilde\psi]$, where 
$\psi_i=\langle h_i \rangle$ are 
the expectation values of the fields $h_i$ in presence of the external sources $j,\tilde{j}$,
 is given by the Legendre transform of  $\log {\cal Z}_k$
(up to the term proportional to $R_k$) \cite{berges02}:
\begin{equation}
\Gamma_k[\psi,\tilde\psi] +\log {\cal Z}_k[j,\tj] = 
\int\! j_i \psi_i -\frac{1}{2} \int_{\bf q}\psi_i\, [R_k]_{ij}\, \psi_{j}
\label{legendre}
\end{equation}
From $\Gamma_k$, one can derive 2-point correlation functions
\be
[\,\Gamma_k^{(2)}\,]_{i_1 i_2}({\bf x}_1,{\bf x}_2, \psi,\tilde\psi) = 
\frac{\delta^2 \Gamma_k[\psi,\tilde\psi]}{\delta\psi_{i_1}({\bf x}_1)\delta\psi_{i_2}({\bf x}_2)}
\ee
and more generally $n$-point correlation functions that we write, 
for future convenience, 
in a $2\times2$ matrix form as (omitting the dependence on the running scale $k$)
\begin{equation}
\Gamma_{i_3,...,i_n}^{(n)}({\bf x}_1,...,{\bf x}_n,\psi,\tilde\psi) =
\frac{\delta^{n-2} \Gamma^{(2)}({\bf x}_1,{\bf x}_2,\psi,\tilde\psi)}
{\delta\psi_{i_3}({\bf x}_3)...\delta\psi_{i_n}({\bf x}_n)} \;.
\end{equation}
The exact flow for $\Gamma_{k}[\psi,\tilde\psi]$ is given by Wetterich's 
equation \cite{berges02}:
\begin{equation}
\partial_k \Gamma_k \!=\! \frac{1}{2}\, {\rm Tr}\! \int_{\bf q}\! \partial_k R_k \cdot G_k
\;\;{\rm with}\;\; G_k\!=\!\left[\Gamma_k^{(2)}+R_k\right]^{-1}\!.
\label{dkgam}
\end{equation}
When $k$ flows from $\Lambda$ to zero, $\Gamma_k$ interpolates 
between $\Gamma_{k=\Lambda}={\cal S}$  and the full effective action $\Gamma_{k=0}$.
Solving Eq.(\ref{dkgam}) is thus equivalent to solving the model. 
This is however impossible to do exactly 
 because it is a non-linear integral partial differential functional equation. 

Here, as usual, our interest lies in the fixed point structure, 
which is determined by the small momentum sector. However, the signature
nonlinear term of the KPZ problem involves gradients, and this seems to
require the faithful description of the full momentum range \cite{canet05}. 
Indeed, expanding the $\Gamma_k^{(n)}$'s 
in powers of their momenta (the so-called derivative expansion),
yields uncontrolled results at least at low orders \cite{canet05}.
Here, by contrast, we define a strategy
which yields closed flow equations for the correlation functions,
and detail it for the two-point functions  $\Gamma_k^{(2)}$.

We first differentiate Eq.(\ref{dkgam}) twice with respect to the fields
and evaluate it, 
without prejudice for the following, in uniform field configurations 
$\Psi_{\rm u}=\{\psi,\tilde\psi\}_{\rm unif}$:
\begin{eqnarray}
\partial_k [\,\Gamma^{(2)}\,]_{ij}(\bp)\! &=& \! {\rm Tr}\! \int_{\bq} \partial_k R(\bq) \cdot G(\bq) \cdot
\!\bigg(\!\!-\!\frac{1}{2}\, \Gamma^{(4)}_{ij}(\bp,-\bp,\bq) \nonumber\\
&& \hspace{-2.2cm} +\Gamma^{(3)}_{i}(\bp,\bq) \cdot G(\bp+\bq) \cdot
\Gamma^{(3)}_{j}(-\bp,\bp+\bq) \bigg) \cdot G(\bq)
\label{dkgam2}
\end{eqnarray}
where the $k$ and $\Psi_{\rm u}$ dependences have been omitted,
as well as the last argument of the $\Gamma^{(n)}$, thanks to 
translation invariance.
Our approximation closes the exact Eq.~(\ref{dkgam2}) 
by approximating the three- and four-point functions.
To this aim, we make use of the following exact relations, 
valid in uniform field
configurations:
\begin{eqnarray}
&&{\Gamma}^{(m,n)}_{k}({\bf q}_1=0,\{{\bf q}_{i}\},\Psi_{\rm u}) =
 \partial_\psi\Gamma^{(m-1,n)}_{k}(\{{\bf q}_{i}\},\Psi_{\rm u})
\label{idpsi}\ \ \ \ \ \ \ \ \\
&&{\Gamma}^{(m,n)}_{k}\!(\{{\bf q}_{i}\},{\bf q}_{m+n}\!\!=\!0,\!\Psi_{\rm u}) =
\partial_{\tilde\psi}\Gamma^{(m,n-1)}_{k}\!(\{{\bf q}_{i}\},\!\Psi_{\rm u})
\label{idbarpsi}
\end{eqnarray}
where, e.g., $\Gamma^{(2,1)}$ is understood as 
$[\,\Gamma_\psi^{(3)}\,]_{\psi \tilde\psi}$, and more generally
$\Gamma^{(m,n)}$ is the $\Gamma^{(m+n)}$ function involving $m$ (resp. $n$)
derivatives w.r.t. $\psi$ (resp. $\tilde\psi$).
To set one ${\bf q}$ to zero, we use the following facts:
Thanks to the presence of the cut-off function $R_k$, 
the momentum dependence of the correlation functions is smooth 
and the internal momentum $\vec{q}$ in (\ref{dkgam2}) 
is effectively limited to $q=\|\vec{q}\|\lesssim k$.
Setting $q=0$ for the $\Gamma_k^{(3)}$ and $\Gamma_k^{(4)}$ functions
in (\ref{dkgam2}) all along the flow is thus a reasonable approximation 
which becomes exact for any finite external momentum in the $k\to 0$ limit.
An exception, though, must be made for the $\Gamma^{(2,1)}$ function:
its bare momentum dependence $\lambda \vec{q}_1\cdot\vec{q}_2$ 
---which stands for the signature nonlinear term of the KPZ problem---
must be kept and the approximation above only implemented on the other terms of the
function.
In addition, we restrict ourselves to zero external frequencies.
Finally, to be able to use  (\ref{idpsi}) and (\ref{idbarpsi}), we set
the internal frequencies to zero in $\Gamma_k^{(3)}$ and $\Gamma_k^{(4)}$
 in (\ref{dkgam2}) (see below). 
At this point, the $\Gamma_k^{(3)}$ and $\Gamma_k^{(4)}$ are replaced 
by derivatives of the
$\Gamma_k^{(2)}$ and Eq.~(\ref{dkgam2}) is closed. 
In fact, the symmetries impose that $\partial_\psi \Gamma_k^{(n)}=0$, 
because of  (\ref{idpsi}) and of the Ward identity 
${\Gamma}^{(m,n)}_{k}(q_1=0,\omega_1,{\bf q}_2,\dots,{\bf q}_{m+n-1}) = 0$ 
(derived from ${\cal T}_{\rm TG}$ and valid for $m>0$ and $(m,n)\neq (1,1)$
 \cite{TBP}).

Solving the problem at this level (with one field dependence) 
was done in equilibrium theories with great success \cite{BMW}. 
Here, this remains very difficult 
because the KPZ symmetries are not easily maintained along the RG flow. 
Thus, on top of the approximation above, we also perform an expansion
around $\tilde\psi=0$ and keep only the leading terms present at the bare level.
All $\Gamma_k^{(3)}$ and  $\Gamma_k^{(4)}$ then vanish, except 
$\Gamma^{(2,1)}=\lambda \vec{q}_1\cdot\vec{q}_2$.
Using various Ward identities imposed by the KPZ symmetries, we find that
the two-point functions are completely determined.
In particular, ${\cal T}_{\rm G}$ and ${\cal T}_{\rm TG}$ impose
that the bare frequency dependence is {\it not} renormalized,
so that, finally, ${\Gamma}^{(1,1)}_{k}$ and ${\Gamma}^{(0,2)}_{k}$
are parameterized by two arbitrary functions $\gamma_k^{11}(q)$ and
$\gamma_k^{02}(q)$, yielding our Ansatz \cite{TBP}:
\begin{equation}
\left\{
\begin{array}{l c l}
{\Gamma}^{(1,1)}_{k}(\vec{q},\omega,\tpsi) & =& i\omega + q^2 \gamma_k^{11}(q)\\
{\Gamma}^{(0,2)}_{k}(\vec{q},\omega,\tpsi) & =& \gamma_k^{02}(q)\\
{\Gamma}^{(2,0)}_{k}(\vec{q},\omega,\tpsi) & =& -\lambda q^2 \tpsi \;.
\end{array}
\right.
\label{anz}
\end{equation}
Note that, in $d=1$, where the extra time-reversal 
symmetry holds, $\gamma_k^{02}=-(2D_k/\nu_k) \gamma_k^{11}$.
Coming back to the internal frequency $\omega$, 
our Ansatz allows to compute exactly the integrals over $\omega$
(which we find convergent). Thus, the large-$\omega$ domain does not
contribute much, which justifies a posteriori our neglecting of higher-order
terms in $\omega$. In fact, we also set to zero the term 
$i\omega q^2$ in ${\Gamma}^{(1,1)}_{k}$, which is small for the momenta of interest.

To treat efficiently the zero-momentum sector of interest near a fixed point,
we introduce dimensionless and renormalized quantities. All momenta
are measured in units of $k$ (e.g. $\tilde q=q/k$).
At the bare level, $\gamma_{\Lambda}^{11}(q)=\nu$ and $\gamma_{\Lambda}^{02}(q)=-2D$. 
We thus define the dimensionless renormalized functions 
$\tilde\gamma_k^{11}(\tilde{q})=\gamma_k^{11}(q)/\nu_k$
and 
$\tilde\gamma_k^{02}(\tilde{q})=-\frac{1}{2D_k} \gamma_k^{11}(q)$.
The running anomalous dimensions $\etan$ (resp. $\etad$) are defined by
$k\partial_k\nu_k=-\etan(k)\nu_k$ (resp. $k\partial_k D_k=-\etad(k)D_k$)
so that at a fixed point, $\nu_k\sim k^{-\etan^*}$ and  $D_k\sim k^{-\etad^*}$.
The scaling exponents are then expressed as
 $z=2-\etan^*$  and $\chi = (2-d+\etad^*-\etan^*)/2$.
To fix the absolute normalizations of $\tilde\gamma_k^{11}(\tilde{q})$ and $\nu_k$
(resp. $\tilde\gamma_k^{02}(\tilde{q})$ and $D_k$), we set 
$\tilde\gamma_k^{11}(0)=1$ (resp. $\tilde\gamma_k^{02}(0)=1$).
(This choice is dictated by the fact that the Ansatz is designed around 
the small-$q$ sector.)

Inserting the Ansatz into (\ref{dkgam2}), we
checked that $\lambda$ does not flow ($k\partial_k \Gamma_k^{(2,0)}=0$),
in agreement with the Galilean invariance,  and thus
the dimensionless running coupling constant reads
$\tilde{g}_k=k^{d-2}\lambda^2 D_k/\nu_k^3$. Its flow equation is:
\begin{equation}
\partial_s \tilde{g}_k = \tilde{g}_k (d-2+3\etan(k)-\etad(k))\;,
\label{eqg}
\end{equation} 
where $\partial_s=k\partial_k$, 
so that $z+\chi=2$ is enforced at any fixed point with $\tilde{g}^*\neq 0$.
We are finally left with the following flow equations for the two running functions
(omitting the $k$ index, using $\vec{u}=\vec{p}+\vec{q}$, 
$\tilde P=-\tilde p^2 +\tilde q^2 +\tilde u^2$,
$\tilde Q=\tilde p^2 -\tilde q^2 +\tilde u^2$,
$\tilde U=\tilde p^2 +\tilde q^2 -\tilde u^2$):
\begin{eqnarray}
\partial_s \tilde{\gamma}^{02}(\tp) \!\!&=&\!\! (\etad \!+\! \tp\,\p_{\tp})\tilde{\gamma}^{02}(\tp)
\!-\! \tilde{g}\!\! \int_{\vec{\tq}} \frac{\tP^4}{\tu^2} 
\frac{\gb(\tu) \tH(\tq,\tu)}{\tilde{L}(\tq,\tu)}\ \ \ \ \label{eq02} \\
\partial_s \tilde{\gamma}^{11}(\tp)\!\!&=&\!\! (\etan \!+\! \tp\,\p_{\tp})\tilde{\gamma}^{11}(\tp)
+\tilde{g}\! \int_{\vec{\tq}} \frac{\tP^2}{\tilde{L}(\tq,\tu)} \times\nonumber\\
&&\hspace{-1cm}\left(\frac{\tU^2}{\tq^2} \ga(\tu) \tH(\tq,\tu)
- \frac{\tQ^2}{\tu^2} \ga^2(\tq)\,\gb(\tu)\,r_\nu(\tq) \right)
\label{eq11}
\end{eqnarray}
\begin{eqnarray}
{\rm with} \;\;\; \tH(\tq,\tu)&=&\ga(\tq) \big((\ga(\tq)+\ga(\tu)\big) r_D(\tq)
\ \ \ \ \ \ \ \ \ \  \nonumber\\
&& - \gb(\tq) \big(2\ga(\tq)+\ga(\tu)\big) r_\nu(\tq) \\
{\rm and} \;\;\;\;\; \tilde{L}(\tq,\tu)&=&\ga^2(\tq)\ga(\tu)\big(\ga(\tq)+\ga(\tu)\big)^2
\end{eqnarray}
where
$r_\nu(\tilde{q}) = \etan r(\tilde{q}) + \tilde{q} r'(\tilde{q})$, 
$\tilde{g}_{11}(\tilde{q}) =\tilde{q}^2 (\tilde\gamma^{11}(\tilde{q})+r(\tilde{q}))$
and similar definitions for $r_D$ and $\tilde{g}_{02}$.
(Note that in $d=1$, the $\tilde\gamma^{02}$ 
and $\tilde\gamma^{11}$ functions are identical.)

\begin{figure}[tp]
\epsfxsize=8cm
\epsfbox{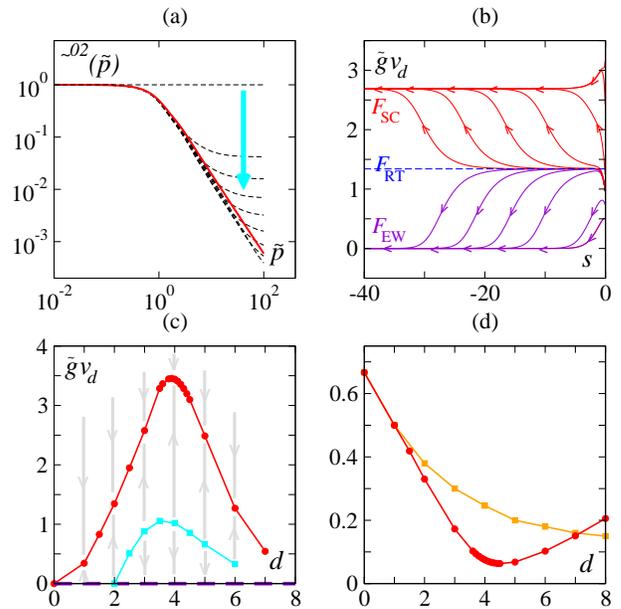}
\caption{(Color online) 
(a) Flow of the $\tilde\gamma^{02}(\tilde p)$ function from its bare shape 
($\tilde\gamma^{02}(\tilde p)=1$) to the KPZ strong coupling fixed point $F_{\rm SC}$ 
(solid red line) in $d=3$.
(b) Flow of $\tilde g v_d$ from various initial bare values in $d=3$.
(The normalization constant $v_d^{-1}=2^{d+1}\pi^{d/2}\Gamma[\frac{d}{2}]$ is related to the integration volume). Flow lines in red (resp. purple) converge to  $F_{\rm SC}$
(resp. $F_{\rm EW}$). The unstable fixed point $F_{\rm RT}$ is given by the blue 
dashed line.
(c) Flow diagram within our approximation
in the $(d,\tilde g v_d)$ plane. 
Red circles: renormalized value $\tilde{g}_{\rm SC}^*$ at $F_{\rm SC}$.
Dashed purple line: Gaussian $F_{\rm EW}$ fixed point.
Cyan squares: bare value $\tilde{g}_{\rm c}$ 
separating the basins of attraction of $F_{\rm SC}$ and  $F_{\rm EW}$. Grey lines
symbolize flow lines.
(d) Variation with $d$ of $\chi=2-z$ for $F_{\rm SC}$
(red circles: our results; orange squares: 
numerical values from \cite{tang92,castellano98}. See also Table~\ref{table}.)}
\label{fig}
\end{figure}

\begin{table}[tp]
\caption{\label{table} Exponent values in integer dimensions. The average numerical values are
extracted from \cite{tang92}. To our knowledge, no estimates of $\omega$ are 
available in the literature.}
\begin{ruledtabular}
\begin{tabular}{lcccc}
  $d$  &   1   &   2   &   3   &   4   \\ \hline
 $\chi$ (this work) & 0.50 & 0.33 & 0.17 & 0.075 \\
 $\chi$ (numerics) &  0.50 & 0.38 & 0.30 & 0.24 \\
$\omega$ (this work) & 0.817 & 0.70 & 0.63 & 0.54 
\end{tabular}
\end{ruledtabular}
\end{table}

We have performed the numerical integration of Eqs.~(\ref{eqg},\ref{eq02},\ref{eq11})
(together with the renormalization conditions 
$\tilde\gamma^{02}(0)=\tilde\gamma^{11}(0)=1$) 
by discretizing momentum on a $\sqrt{\tilde p}$ mesh of typical spacing $5.10^{-2}$, 
using 5-point finite-difference expressions for the 
$\tilde p\partial_{\tilde p}$ terms,
Euler explicit time-stepping with a typical time step  $\Delta s =-10^{-4}$. 
All integrals were estimated using Simpson's rule and Cartesian coordinates. 
We have checked the robustness of the results presented below
against numerical resolution and against variations of the momentum range
used for the cut-off function $r$ (typically 4 in units of $\tilde p$). In short, 
the simplest numerical techniques performed nicely, with typical runs taking minutes on 
a current computer.

We systematically followed the flow of functions $\tilde\gamma^{02}$ and
$\tilde\gamma^{11}$, together with the running coupling constant $\tilde g$ and 
running exponents $\eta_D$ and $\eta_{\nu}$ from the bare initial condition
$\tilde\gamma^{02}(\tilde p)=\tilde\gamma^{11}(\tilde p)=1$ down to $k=0$ ($s\to -\infty$). 
Figure~\ref{fig}c summarizes the flow diagram obtained in our approximation.
In all dimensions studied (i.e. up to $d=8$), we have found,
besides the Edwards-Wilkinson fixed point $F_{\rm EW}$,
a fully-attractive non-trivial fixed point $F_{\rm SC}$.
In all $d$, we find generic scaling,  {\it i.e.} the flow always 
reaches one of these fixed points.
For $d<2$, $F_{\rm SC}$ is reached from any initial condition.
Along the flow, the $\tilde\gamma^{02}(\tilde p)$ and
$\tilde\gamma^{11}(\tilde p)$ functions gently deform to reach a fixed shape with
an algebraic tail at large $\tilde p$ (Fig.~\ref{fig}a).
For $d>2$, $F_{\rm EW}$
becomes locally fully attractive, and
there exists a critical 
bare value $\tilde g_{\rm c}$ separating the basins of attraction of $F_{\rm SC}$
and $F_{\rm EW}$ (Fig.~\ref{fig}b).
Right at  $\tilde g_{\rm c}$ and for $d<5$, the flow reaches an 
unstable fixed point $F_{\rm RT}$ which drives the roughening transition. 
In $d=2$, $F_{\rm RT}$ coincides with $F_{\rm EW}$, 
and becomes non-Gaussian for larger dimensions. 
Because Galilean symmetry is respected (and its coupling constant 
${\tilde g}^*_{\rm SC}$ is non zero), $F_{\rm SC}$ is characterized by a single exponent.
We thus only discuss below the values obtained for $\chi$ \cite{NOTE}.
Our work also gives access to the subleading exponent $\omega$ characterizing
the approach to $F_{\rm SC}$. 
Table~\ref{table} and Figure~\ref{fig}d contain our estimates. 
For $d\lesssim 4$, $\chi$ decreases almost linearly with $d$,
with the exact value $\frac{1}{2}$ (resp. $\frac{2}{3}$) recovered in $d=1$
(resp. $d=0$), and a reasonable but deteriorating
agreement with numerical values for $2\le d\le 4$. In higher dimensions,
$\chi$ increases with $d$, at odds with both numerical values and 
the scenario of $d=4$
being an upper critical dimension beyond which $\chi=0$
\cite{lassig97,colaiori01,fogedby05}.
Regarding $F_{\rm RT}$, we record negative values of $\chi$ for $2<d<5$,
which is reminiscent of perturbative results performed at fixed $d$ \cite{frey94}
but in contradiction with exact results dictating $\chi=0$ \cite{doty,wiese98}.
For $d>5$,  $F_{\rm RT}$ seems unreachable from bare 
initial conditions, probably an
effect of the crudeness of our approximation.

Several observations hint at a qualitative change occuring around $d=4$: 
the values of ${\tilde g}^*_{\rm SC}$ and $\tilde g_{\rm c}$  first increase, 
then decrease (Fig.~\ref{fig}c). 
For large $d$, the unstable fixed point $F_{\rm RT}$ becomes ill-defined.
At this stage, we are unable to decide whether the change of behavior near $d=4$
is related to the existence of an upper critical dimension or whether
 this is just an effect of the deterioration of the 
quality of our scheme in large $d$.
Higher orders of our approximation, if only by adding
the next terms in $\tilde\psi$ to the two-point functions, might yield
the answer to this important question. 
At any rate, we believe our results represent a breakthrough,
essentially because our approximation scheme is able to deal with 
correlation functions in a non-perturbative and systematically improvable way.
In particular, as opposed to the mode-coupling approximation, which
assumes scaling forms for the two-point functions, our equations flow
generically to a fully-stable strong-coupling fixed point 
---which roots the existence of generic scaling---,
or to the EW fixed point for $d>2$, yielding, 
for the first time, the correct phase diagram of the KPZ equation
within a RG approach.

\end{document}